\documentclass[sigconf]{acmart}
\AtBeginDocument{%
  }

\acmConference[PPoPP'27]{Make sure to enter the correct
  conference title from your rights confirmation email}{March 20--24,
  2027}{Salt Lake City, Utah}

\usepackage{amsmath,amsfonts}
\usepackage{graphicx}
\usepackage{textcomp}
\usepackage{xcolor}
\usepackage{xspace}
\usepackage{float}
\hypersetup{colorlinks,linkcolor=purple,citecolor=purple}
\usepackage[linesnumbered,ruled,vlined]{algorithm2e}
\usepackage{booktabs}
\usepackage{pifont} 

\newcommand{\inc}[1]{\textbf{(+\,#1)}}

\definecolor{revblue}{RGB}{45,110,210}
\definecolor{revblack}{RGB}{0,0,0}
\newcommand{\rev}[1]{{\color{revblack}#1}}

\newcommand{\ALGtag}[2]{%
\makebox[3em][l]{\textcolor{#1}{#2}}%
}

\newcommand{\sysname}{EvoCG\xspace}
\begin{document}
\settopmatter{printacmref=false}    
\setcopyright{none}                 

\title{\sysname: Self-Evolving Precision Control for Faster Conjugate Gradient}


\author{Jiaqi Zhai}
\email{jqzhai@hust.edu.cn}
\orcid{0000-0002-9767-812X}
\authornotemark[1]
\authornotemark[2]
\authornotemark[3]
\affiliation{%
  \institution{Huazhong University of Science and Technology}
  \city{Wuhan}
  \country{China}
}

\author{Haozhi Han}
\email{haozhi.han@stu.pku.edu.cn}
\authornote{Both authors contributed equally to this research.}
\authornote{Work done during an internship at Institute for AI Industry Research(AIR), Tsinghua University.}
\affiliation{%
  \institution{School of Computer Science, Peking University}
  \city{Beijing}
  \country{China}
}

\author{Kaiyi Huang}
\email{huangky@hust.edu.cn}
\authornote{National Engineering Research Center for Big Data Technology and System; Services Computing Technology and System Lab; Cluster and Grid Computing Lab; School of Computer Science and Technology, Huazhong University of Science and Technology, Wuhan 430074, China.}
\affiliation{%
  \institution{Huazhong University of Science and Technology}
  \city{Wuhan}
  \country{China}
}

\author{Wenju Zhao}
\email{wjzh@hust.edu.cn}
\authornotemark[3]
\affiliation{%
  \institution{Huazhong University of Science and Technology}
  \city{Wuhan}
  \country{China}
}

\author{Yifeng Chen}
\email{cyf@pku.edu.cn}
\affiliation{%
  \institution{School of Computer Science, Peking University}
  \city{Beijing}
  \country{China}
}

\author{Yunquan Zhang}
\email{zyq@ict.ac.cn}
\affiliation{%
  \institution{Institute of Computing Technology, Chinese Academy of Sciences}
  \city{Beijing}
  \country{China}
}

\author{Ting Cao}
\email{tingcao@mail.tsinghua.edu.cn}
\affiliation{%
  \institution{Institute for AI Industry Research, Tsinghua University}
  \city{Beijing}
  \country{China}
}

\author{Yunxin Liu}
\email{liuyunxin@air.tsinghua.edu.cn}
\affiliation{%
  \institution{Institute for AI Industry Research, Tsinghua University}
  \city{Beijing}
  \country{China}
}

\author{Hai Jin}
\email{hjin@hust.edu.cn}
\authornotemark[3]
\affiliation{%
  \institution{Huazhong University of Science and Technology}
  \city{Wuhan}
  \country{China}
}

\author{Xuanhua Shi}
\email{xhshi@hust.edu.cn}
\correspondingauthor
\authornotemark[3]
\affiliation{%
  \institution{Huazhong University of Science and Technology}
  \city{Wuhan}
  \country{China}
}

\author{Kun Li}
\email{likungw@gmail.com}
\correspondingauthor
\affiliation{%
  \institution{Institute for AI Industry Research, Tsinghua University}
  \city{Beijing}
  \country{China}
}

\begin{abstract}
Modern GPUs offer far higher throughput in low precision than in FP64, but exploiting this in conjugate gradient (CG) solvers is difficult because tolerance to reduced precision evolves with the Krylov state. We present \sysname, a mixed-precision CG solver that treats tile-wise precision as an online control action rather than a fixed setting: \rev{a reinforcement-learning policy in the iteration loop selects per-tile precision from the evolving solver state}, made profitable by co-design with a Tile-CSC unified mixed-precision SpMV kernel and a low-overhead in-loop runtime. Across 47 SPD matrices from eight scientific domains, \sysname is the fastest solver overall, with geometric-mean speedups of 2.52$\times$, 3.06$\times$, 2.48$\times$, 1.93$\times$, and 6.99$\times$ over cuSPARSE, PETSc, Ginkgo, Mille-FP64, and Mille-Mixed. \rev{On the same substrate, it improves time-to-solution over \sysname-FP64 by 1.28$\times$ at only 11.4\% more iterations and 1.09\% overhead, with FP64-recomputed final true-residual norms nearly identical to those of FP64, and extends to Jacobi-preconditioned CG without retraining, yielding a further 1.15$\times$ geometric-mean speedup over \sysname-FP64.}
\end{abstract}






\renewcommand{\shortauthors}{Zhai et al.}

\maketitle

\section{Introduction}

The rise of AI has reshaped the precision landscape of modern processors~\cite{gupta2015deep, micikevicius2018mixed}: contemporary GPUs deliver only tens of TFLOPS in FP64, whereas lower-precision formats such as FP32, TF32, FP16, BF16, and INT8 offer peak throughputs one to two orders of magnitude higher~\cite{nvidia_tensor_core, choquette2023nvidia}. Scientific computing, in contrast, still relies predominantly on uniform double precision for robustness and convergence guarantees~\cite{dongarra2024hardware}, creating a fundamental efficiency gap: massive low-precision capacity that traditional HPC workloads cannot exploit safely and systematically~\cite{vieuble2022mixed}.

This mismatch is pronounced in Krylov subspace solvers~\cite{golub1997iterative,hestenes1952methods,saad2003iterative}. Among them, the Conjugate Gradient (CG) method is a cornerstone HPC sparse solver underpinning finite element analysis~\cite{chen2002application}, computational fluid dynamics~\cite{cheng2014cfd}, and computational electromagnetics~\cite{vargas2022subspace}. \rev{Yet production CG implementations—including cuSPARSE-based solvers~\cite{cusparse}, PETSc~\cite{petsc}, and Ginkgo~\cite{anzt2022ginkgo}—still execute the entire iteration uniformly in FP64, leaving the abundant low-precision throughput of modern GPUs untapped.} 

Within the CG iteration, Sparse Matrix--Vector Multiplication (SpMV) dominates runtime, often 60\%--95\% of time-to-solution for large sparse systems~\cite{bell2009implementing, williams2009roofline, liu2015csr5}. Lowering SpMV precision is thus a major lever for accelerating CG. Our experiments show that, on average, 56.9\% of SpMV computation across the 47 matrices is executed in FP32 or BF16. The challenge is that reduced-precision SpMV perturbs the Krylov recurrence~\cite{simoncini2003theory, bouras2005inexact}, and CG's tolerance to such perturbations evolves with the solver state, potentially slowing or even destabilizing convergence~\cite{carson2018accelerating, higham2022mixed, yang2024mille, chen2025mixed}.

Mixed-precision CG has been explored in several forms, yet precision remains largely an open-loop decision: early approaches confine low precision to preconditioners or inner solvers, keeping SpMV in high precision~\cite{suzuki2025nested, ina2023new}.
Recent GPU-oriented methods lower SpMV precision but follow static or rule-based schedules agnostic to the Krylov state~\cite{yang2024mille}\rev{---schedules that, in our evaluation, frequently fail to converge on sensitive problems (Sec. \ref{subsec:whyLearnedPC}).}
RL-based approaches select precision as coarse-grained offline autotuning, \rev{with no execution substrate to turn low precision into actual speedup}~\cite{chen2025mixed}. Existing methods thus expose precision heterogeneity without closing the feedback loop \rev{or delivering its performance benefit}.

These limitations point to a deeper question:
\emph{Can Krylov solvers become self-evolving systems in which precision is adaptively controlled throughout the iteration process?}

We answer this question with \sysname, built on a key insight: \textbf{precision is control, not configuration}---not fixed offline as a static setting, but regulated online as a solver action driven by the evolving Krylov state. \sysname realizes this as a closed loop: a reinforcement-learning policy embedded in the CG iteration reads the Krylov state and selects the tile-wise precision, whose induced inexactness in turn shapes the next state. Precision thus becomes a controllable input, and inexactness a managed degree of freedom that balances convergence and hardware efficiency.

\rev{Turning low precision into end-to-end solver speedup requires meeting three requirements simultaneously: the solver must \emph{recognize} when reduced precision is tolerable for the current Krylov state, \emph{convert} that tolerance into real speedup despite tile-level precision heterogeneity, and \emph{integrate} this control at overhead low enough for an iterative solver---missing any one forfeits the benefit.} \textbf{\sysname} addresses these with three co-designed modules: \textbf{EvoCtrl} maps the current search direction $p_k$ to a tile-level numerical state and selects a tile-wise precision assignment over \{FP64, FP32, BF16\} with a learned policy; \textbf{EvoSpMV} executes the resulting heterogeneous SpMV in a single warp-centric kernel built on Tile-CSC, a storage format co-designed to absorb tile-level precision heterogeneity without control-flow divergence; and \textbf{EvoRun} amortizes the control overhead on the critical path through device-side inference, interval-based reuse, and CUDA-Graph replay.

On 47 SPD SuiteSparse matrices spanning eight scientific domains, \sysname\ is the fastest solver overall, with geometric-mean speedups of 2.52$\times$, 3.06$\times$, 2.48$\times$, 1.93$\times$, and 6.99$\times$ over cuSPARSE, PETSc, Ginkgo, Mille-FP64, and Mille-Mixed. Against \sysname-FP64, our FP64 CG solver on the identical substrate, dynamic precision control improves end-to-end performance by 1.28$\times$ on average at only 11.4\% more iterations, and the average policy overhead is only 1.09\%. \rev{The same policy further extends to Jacobi-preconditioned CG without retraining, delivering a 1.15$\times$ geometric-mean speedup over \sysname-FP64.}

This paper makes the following contributions:
\begin{itemize}
    \item We exploit CG's state-dependent precision tolerance to formulate mixed-precision CG as a reinforcement-learning control problem, turning tile-wise precision into a learned solver action that yields end-to-end speedup.
    \item We turn tile-wise low-precision decisions into solver-level speedup with Tile-CSC and a unified warp-centric SpMV kernel that runs mixed-precision tiles without control-flow divergence.
    \item We keep precision control on the CG critical path at low overhead, via device-side inference, interval-based reuse, and CUDA-Graph-replayed fused execution.
\end{itemize}

\section{Background and Challenges}
\label{sec:background}

\subsection{Conjugate Gradient}
\label{subsec:cg}

The Conjugate Gradient (CG) method solves symmetric positive definite linear systems $Ax=b$ by iteratively constructing Krylov subspaces. At iteration $k$, CG maintains a solver state
\begin{equation}
\label{eq:solver_state}
z_k=\{x_k,r_k,p_k\},
\end{equation}
where $x_k$ is the approximate solution, $r_k$ the residual, and $p_k$ the search direction.

\begin{figure}[b]
    \centering
    \includegraphics[width=1\linewidth]{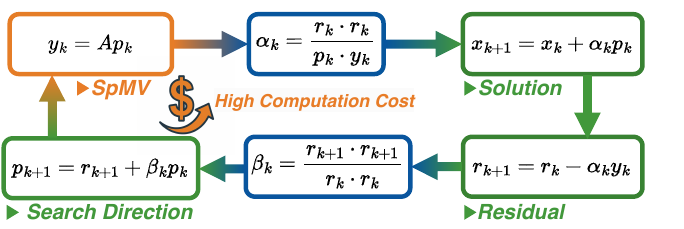}
    \caption{One iteration of the CG recurrence. SpMV is the dominant cost.}
    \label{fig:cg_iteration}
\end{figure}

CG evolves this state through the short recurrence shown in Fig.~\ref{fig:cg_iteration}, consisting of one SpMV, two dot products, and three AXPY operations per iteration.

\begin{figure}[h]
\centering 
\includegraphics[width=1.0\linewidth]{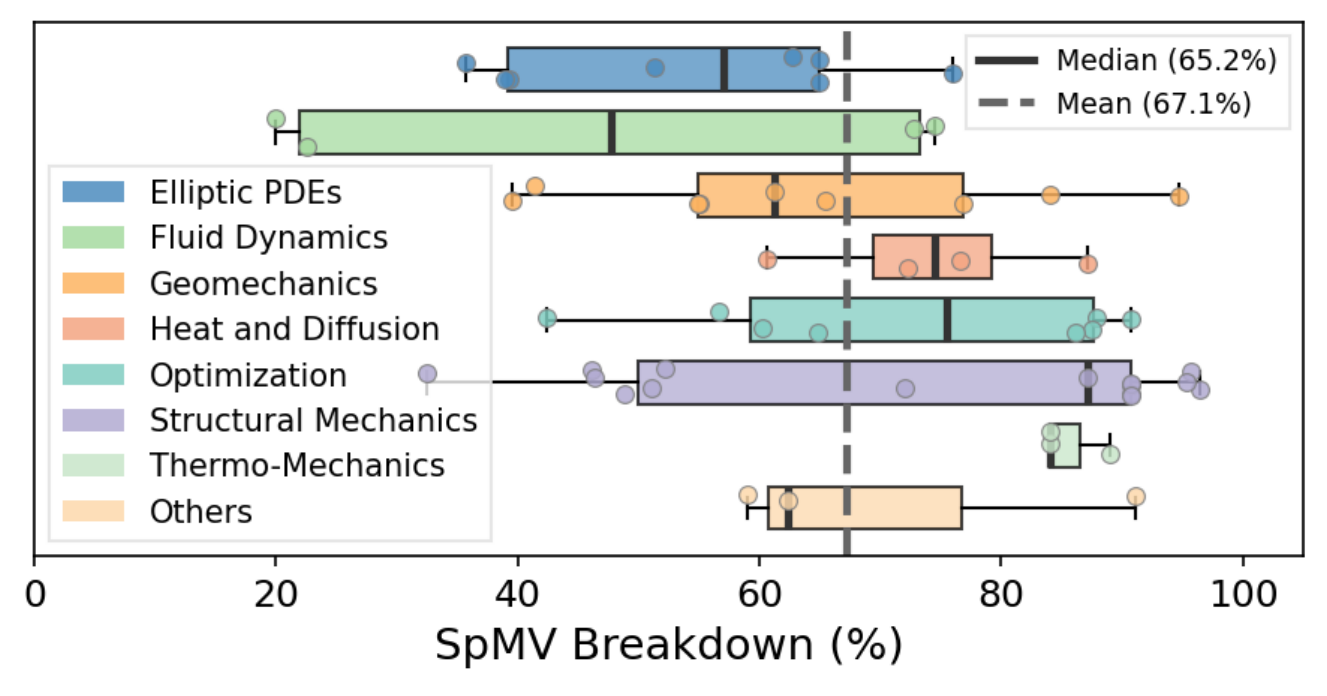} 
\caption{Dominance of SpMV in CG execution across 47 sparse linear systems from eight scientific domains.} 
\label{fig:spmv_breakdown} 
\end{figure}

Among these operations, SpMV dominates execution time: Fig.~\ref{fig:spmv_breakdown} shows that across 47 sparse systems from eight domains, it accounts for 67.1\% of runtime on average and exceeds 80\% for many large problems. Reducing SpMV precision improves arithmetic efficiency but introduces numerical perturbations into the CG recurrence:

\begin{equation}
\tilde{y}_k=(A+E_k)p_k,
\end{equation}

where $E_k$ represents the error introduced by reduced-precision arithmetic. Inexact Krylov theory interprets this as applying CG to a sequence of perturbed operators and shows that such perturbations affect both attainable accuracy and convergence behavior~\cite{simoncini2003theory,bouras2005inexact,van2004inexact}.

The impact on attainable accuracy is characterized by the residual gap between the true residual $r_m=b-A\tilde{x}_m$ and the recursively updated residual $\tilde r_m$:

\begin{equation}
\|r_m-\tilde r_m\|
\leq
\sum_{k=1}^{m}\|E_k\||\eta_k^{(m)}|,
\end{equation}

where $\eta_k^{(m)}$ represents the contribution of the $k$-th perturbation to the residual gap~\cite{simoncini2003theory}. Therefore, controlled perturbations can preserve the attainable final accuracy.

For short-recurrence methods such as CG, perturbations may further alter the convergence trajectory through residual drift and loss of orthogonality~\cite{van2004inexact,higham2022mixed,greenbaum1989behavior}, introducing a fundamental tradeoff: lower precision reduces the cost of each SpMV, but the added perturbations may increase the total iteration count.

Inexact Krylov theory characterizes the relationship between perturbations and convergence, but it does not determine which precision strategy minimizes end-to-end solution time. This motivates adaptive precision control, where each precision decision must balance immediate execution cost and future convergence behavior.

\subsection{Reinforcement Learning}
\label{subsec:rl}

The above precision-control problem naturally forms a sequential decision process. At each CG iteration, a controller observes the evolving solver state, selects precision actions, and receives feedback from both execution cost and convergence behavior. Unlike static precision assignment or residual-only heuristics, the optimal decision depends on long-term effects across iterations.

Reinforcement learning provides a natural framework for this setting by optimizing policies through long-term rewards. Proximal Policy Optimization (PPO)~\cite{schulman2017ppo} is adopted because its clipped policy updates improve optimization stability for long-horizon control problems~\cite{xie2024simple}.

\subsection{Precision Control Formulation}
\label{subsec:precision_control}




CG maps naturally onto an MDP. The evolving solver state $\{x_k,r_k,p_k\}$ supplies a Markov state, tile-wise precision forms a discrete action space, and the CG recurrence itself is the transition dynamics, requiring no learned environment model. The one non-obvious element is the reward, which must encode a delayed objective: a cheap low-precision SpMV now may cost extra iterations later, so end-to-end time-to-solution, not per-iteration cost, is what the policy must optimize. Sec.~\ref{sec:method_control} instantiates each component.

\subsection{Challenges}

Unlocking low precision in CG while preserving convergence robustness poses three challenges.

\textbf{C1: Capturing long-term numerical effects of precision decisions.}
Precision choices affect not only current SpMV cost but also future convergence through the Krylov recurrence, making it difficult to design states and rewards that capture the accuracy-performance tradeoff.

\textbf{C2: Performance-effective execution under sparsity and heterogeneous precision.}
Fine-grained precision control introduces irregular computation, whereas GPUs favor regular execution, requiring cross-layer co-design of sparse format, kernel mapping, and scheduling.

\textbf{C3: Low-overhead integration of adaptive control.}
A learned precision policy must be integrated into the CG execution loop with minimal runtime overhead, requiring efficient inference and solver-aware execution mechanisms.
\section{\sysname}

\subsection{Overview}

\begin{figure}[b]
    \centering
    \includegraphics[width=0.9\linewidth]{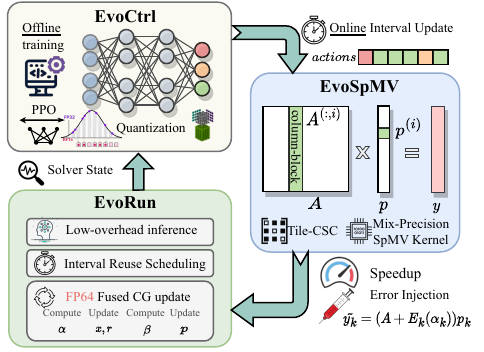}
    \caption{\sysname overview.}
    \label{fig:system_overview}
\end{figure}

As shown in Fig.~\ref{fig:system_overview}, \sysname comprises three tightly coupled modules: EvoCtrl (control) learns an RL policy that selects tile-wise precision from the evolving Krylov state. EvoSpMV (execution) turns these decisions into an efficient sparse substrate via the co-design of Tile-CSC and a unified warp-centric mixed-precision kernel. EvoRun (runtime) embeds policy execution into the CG loop via low-overhead inference, interval reuse, and fused execution. Together, they determine what precision to use, how to execute it efficiently, and when this control enters the solver.

\subsection{EvoCtrl: Closed-Loop Precision Control}
\label{sec:method_control}

\subsubsection{Control Objective Formulation}

EvoCtrl minimizes CG time-to-solution while preserving convergence robustness. Let $T(\{a_k\})$ be the total execution time to reach a prescribed residual tolerance under precision decisions $\{a_k\}$:
\begin{equation}
\label{eq:global_objective}
T(\{a_k\}) = \sum_{k=0}^{K(\{a_k\})-1} C(a_k),
\end{equation}
where $a_k$ is the precision configuration at iteration $k$, $C(a_k)$ is its per-iteration cost, and $K(\{a_k\})$ is the iteration count.

Making this dependence explicit, mixed-precision SpMV is modeled as:
\begin{equation}
\label{eq:perturbed_spmv_action}
\tilde{y}_k = (A + E_k(a_k)) p_k,
\end{equation}
where $E_k(a_k)$ denotes the perturbation induced by action $a_k$. This yields the controlled solver dynamics:
\begin{equation}
\label{eq:controlled_dynamics}
z_{k+1} = F(z_k, a_k),
\end{equation}
where $z_k$ denotes the solver state. Precision decisions therefore affect both the immediate execution cost and the future convergence trajectory, making EvoCtrl recurrence-driven control rather than a static configuration rule.

\subsubsection{Precision Control MDP}

\sysname models the controlled CG process as an MDP $(\mathcal{S}, \mathcal{A}, \mathcal{P}, \mathcal{R}, \gamma)$: each CG iteration is a decision step and a complete solve an episode.

\paragraph{\textbf{State Space $\mathcal{S}$}}
\label{state_space}

The MDP state is constructed from the CG search direction $p_k$, motivated by two properties. First, through the CG recurrence,
\begin{equation}
p_{k+1} = r_{k+1} + \beta_k p_k,
\end{equation}
$p_k$ carries accumulated information about the convergence regime. Second, $p_k$ is the direct input to SpMV, so its tile-level distribution determines the magnitude of partial sums and their sensitivity to reduced precision. These two properties make $p_k$ a well-suited state for convergence-aware control.

To expose both global solver dynamics and localized computation, $p_k \in \mathbb{R}^n$ is partitioned into $n_{\text{tiles}}$ segments of size $C$, each normalized to form the policy input:
\begin{equation}
\label{eq:state_mapping}
\left.
\begin{aligned}
p_k^{(i)} &= p_k[iC : (i+1)C-1],\\
\phi(p_k^{(i)}) &= p_k^{(i)}/\max(\|p_k^{(i)}\|_2,\delta),\\
s_k &= (\phi(p_k^{(0)}),\dots,\phi(p_k^{(n_{\text{tiles}}-1)})),
\end{aligned}
\right.
\end{equation}
where $\delta$ prevents division by zero.

\paragraph{\textbf{Action Space $\mathcal{A}$}}

At iteration $k$, EvoCtrl outputs a tile-wise precision assignment vector:
\begin{equation}
\label{eq:action_definition}
a_k = (a_k^{(0)}, \dots, a_k^{(n_{\text{tiles}}-1)}),
\end{equation}

where $a_k^{(i)} \in \{0,1,2\}$ encodes the precision \{FP64, FP32, BF16\} for the vector tile $p_k^{(i)}$ and its associated column block, as defined in Eq.~(\ref{eq:SpM_cols_block}):

\begin{equation}
\label{eq:SpM_cols_block}
A^{(:,i)} = A[:,\, iC : (i+1)C - 1],
\end{equation}




\paragraph{\textbf{Reward Function $\mathcal{R}$}}

The reward is designed as an objective for minimizing end-to-end time-to-solution. It jointly captures convergence progress, execution cost, and successful termination:

\begin{equation}
\label{eq:reward}
R_k =
w_1 \ln \frac{\|r_{k-1}\|}{\|r_k\|}
- w_2 \frac{C(a_k)}{n_{\text{tiles}}}
+ w_3 \mathbb{1}\left(\frac{\|r_k\|}{\|r_0\|}\leq\tau\right),
\end{equation}

where $\tau$ is the convergence tolerance. The first term rewards residual reduction, its logarithmic form normalizing improvement across iterations. The second term penalizes per-iteration cost $C(a_k)$ through a scale-independent per-tile signal, letting the same reward generalize across matrix scales. The terminal bonus prevents aggressive low-precision choices that cut immediate cost but miss the target accuracy. We set $w_1=10$, $w_2=0.2$, and $w_3=30$, chosen from stability-oriented defaults and reused unchanged across all 47 matrices, with the convergence terms weighted higher because early low-precision exploration can destabilize CG.

\paragraph{\textbf{Transition Dynamics $\mathcal{P}$ and Discount Factor $\gamma$}}
The transition dynamics are the CG recurrence itself, so PPO interacts directly with the mixed-precision solver as the environment without an explicit model of $\mathcal{P}$. We set $\gamma = 0.99$.

\subsubsection{Policy Training}
\label{traing_methodology}

EvoCtrl is instantiated as a lightweight MLP parameterizing both policy and value: each normalized segment $\phi(p_k^{(i)}) \in \mathbb{R}^{C}$ passes through a two-layer ReLU backbone. The policy head outputs a three-class distribution over \{FP64, FP32, BF16\} yielding $a_k^{(i)}$, and the value head predicts the expected return.


The network is trained with PPO in a solver-in-the-loop manner, summarized in Algorithm~\ref{alg:ppo_cg}\rev{: lines~\ref{algline:tileloop}--\ref{algline:action} construct tile states and sample tile-wise actions, line~\ref{algline:spmv} executes the mixed-precision SpMV inside the CG recurrence, the SOVR lines perform the standard CG updates, line~\ref{algline:reward} evaluates the reward of Eq.~(\ref{eq:reward}), and line~\ref{algline:ppo} stores solver transitions and apply the clipped
PPO update whenever 1024 transitions have been collected.}
The tile-wise precision actions learned by EvoCtrl are later consumed by EvoSpMV, which serves as the sparse execution substrate for efficient GPU realization.

\begin{algorithm}[tb]
\small
\caption{\sysname Training Phase.\\
\parbox[t]{0.85\linewidth}{
\footnotesize
\textcolor{teal}{INIT}: policy setup and FP64 baseline estimation.\\
\textcolor{blue}{ENV}: RL controller operations.\\
\textcolor{orange}{SOVR}: CG recurrence updates.\\
\textcolor{purple}{SpMV}: tile-wise mixed-precision SpMV execution.
}}
\label{alg:ppo_cg}
\ALGtag{blue}{ENV}\KwIn{Matrix $A$, right-hand side $b$, tile size $C$}
\ALGtag{blue}{ENV}\KwOut{Trained policy network $\pi_\theta$}

\ALGtag{teal}{INIT}Initialize policy $\pi_\theta$, value network $V_\theta$\;
\ALGtag{teal}{INIT}Run full FP64 CG to obtain $base\_iteration$\;
\ALGtag{blue}{ENV}$max\_iter \gets 1.5 \times base\_iteration$\;
\ALGtag{blue}{ENV}$n_{tiles} \gets (\left\lceil A.\mathrm{shape}[1]/C\right\rceil);$

\For{$episode = 1$ \textbf{to} $1000$}{

    \ALGtag{orange}{SOVR}Initialize $x \gets 0$, $r \gets b - A x$, $p \gets r$\;

    \For{$k=0$ \textbf{until} $\|r\|/\|r_0\| \le \tau$ \textbf{or} $k=\text{max\_iter}$}{

        \For{$i \gets 0$ \textbf{to} $n_{\text{tiles}}-1$}{\label{algline:tileloop}
        \ALGtag{blue}{ENV}$p^{(i)} \gets p[iC : (i+1)C-1]$\;
        \ALGtag{blue}{ENV}$s^{(i)} \gets \phi(p^{(i)})$\;
        \ALGtag{blue}{ENV}$a^{(i)} \sim \pi_\theta(s^{(i)})$\;\label{algline:action}
        }
        \For{each tile $i$}{    
            \ALGtag{purple}{SpMV}$y \gets y + A^{(:,i)} p^{(i)}$ \quad (by $a^{(i)}$)\;\label{algline:spmv}
        }

        \ALGtag{orange}{SOVR}$\alpha \gets (r^T r)/(p^T y)$\;
        \ALGtag{orange}{SOVR}$x \gets x + \alpha p$\;
        \ALGtag{orange}{SOVR}$r_{\text{new}} \gets r - \alpha y$\;
        \ALGtag{orange}{SOVR}$\beta \gets (r_{\text{new}}^T r_{\text{new}})/(r^T r)$\;
        \ALGtag{orange}{SOVR}$p \gets r_{\text{new}} + \beta p$\;
        \ALGtag{blue}{ENV}$R \gets
        w_1 \ln \frac{\|r\|}{\|r_{\text{new}}\|}
        - w_2 \frac{C(a)}{n_{\text{tiles}}}
        + w_3 \, \mathbb{1}\!\left(\frac{\|r_{\text{new}}\|}{\|r_0\|} \le \tau \right)$\;\label{algline:reward}
        \ALGtag{orange}{SOVR}$r \gets r_{\text{new}}$\;

    }
    \ALGtag{blue}{ENV}Update $\pi_\theta$ and $V_\theta$ using PPO objective\;\label{algline:ppo}
}
\ALGtag{blue}{ENV}\KwRet{$\pi_\theta$}
\end{algorithm}

\subsection{EvoSpMV: Mixed-Precision Sparse Execution}

EvoSpMV is the execution module of \sysname. It executes EvoCtrl's tile-wise precision decisions under irregular sparsity without fragmenting into precision-specific kernel paths. The challenge is that mixed precision makes SpMV heterogeneous in both sparsity and arithmetic cost. EvoSpMV addresses this by co-designing the Tile-CSC layout with a unified warp-centric mixed-precision kernel, confining precision heterogeneity to tiles while keeping execution regular at the warp-group and lane levels.

\begin{figure}[h]
    \centering
    \includegraphics[width=1\linewidth]{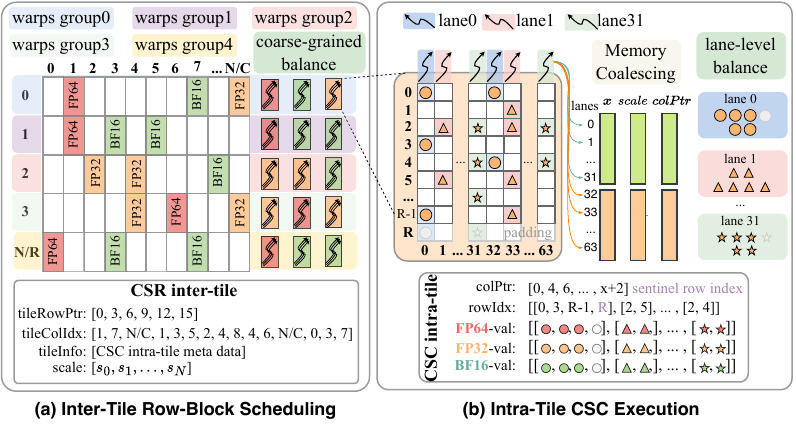}
    \caption{EvoSpMV: co-design of Tile-CSC and a warp-centric mixed-precision SpMV kernel.}
    \label{fig:tileCSC}
\end{figure}

As shown in Fig.~\ref{fig:tileCSC}(a), the sparse matrix is partitioned into $R \times C$ tiles, where $C$ is fixed to 64 and $R$ is determined during preprocessing. Each tile corresponds to the intersection between an $R$-row block and a consecutive 64-column block, serving as the unified granularity for Tile-CSC storage, EvoSpMV scheduling, and runtime precision control.

\subsubsection{Inter-Tile Row-Block Scheduling}

The inter-tile challenge is balancing warp-group workload under heterogeneous tile precisions. \sysname organizes non-empty tiles row-block-wise in CSR style, where \texttt{tileRowPtr} delimits each row-block's tile interval and \texttt{tileColIdx} records column-block indices (Fig.~\ref{fig:tileCSC}(a)). Each row-block is assigned a fixed group of $W$ warps traversing its non-empty tiles in a strided manner (warp $w$ processes tiles $t_0+w,\, t_0+w+W,\, \ldots$). 

This CSR-style organization disperses tiles of different precision across warp groups: each group spans a mixture of FP64, FP32, and BF16 tiles rather than a precision-homogeneous stream, averaging the precision-dependent cost variation into a uniform workload. A CSC-style layout would instead cluster same-precision tiles within a warp group, causing significant inter-warp imbalance.

\subsubsection{Intra-Tile CSC Execution}
\label{subsubsection:unified_precision_path}

The intra-tile challenge is dynamic precision selection without sacrificing warp-level regularity. As Fig.~\ref{fig:tileCSC}(b) shows, EvoSpMV encodes each tile in CSC format and assigns one warp per tile. The tile descriptor \texttt{tileInfo} records the global-memory offsets of \texttt{colPtr}, \texttt{rowIdx}, and the value streams: within a tile, \texttt{colPtr} (length $C+1$) defines each column's nnz range, and \texttt{rowIdx} stores row offsets relative to the row-block base. The warp traverses the 64 columns cooperatively, lane $i$ processing columns $i$ and $i+32$. Accesses to \texttt{x}, \texttt{scale}, and \texttt{colPtr} coalesce across lanes, while \texttt{rowIdx} and value accesses stay column-local, confining sparse irregularity to column-local traversal rather than warp-wide divergence.

\begin{figure*}[t]
    \centering
    \includegraphics[width=1\linewidth]{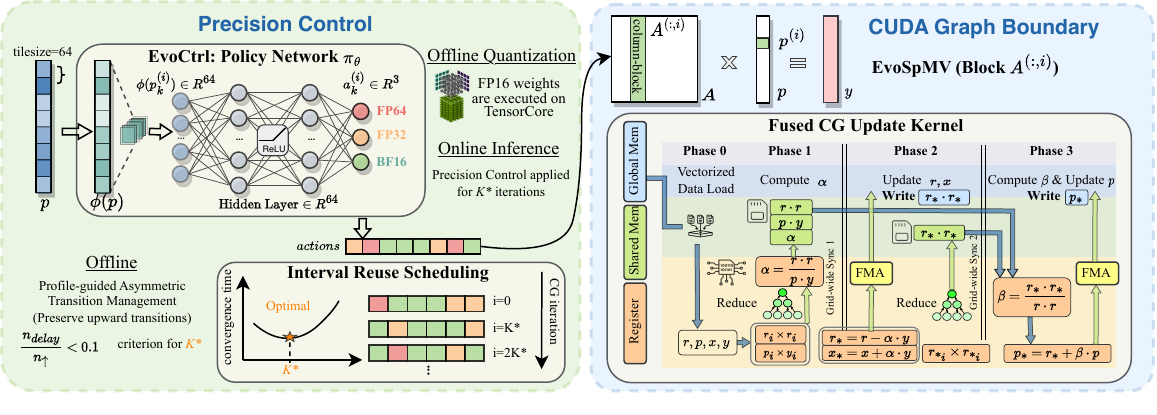}
    \caption{EvoRun: system-level realization of \sysname\ with in-loop policy inference---GPU-resident policy execution, action reuse over $K^\star$ iterations, and CUDA-Graph replay of the fused solver backbone.}
    \label{fig:EvoRun}
\end{figure*}

This column-centric path aligns naturally with the mixed-precision layout. Quantization is applied per column, so all nonzeros in a column share one scaling factor, and column traversal amortizes the loads of $x[j]$ and $s[j]$. Since FP32 and BF16 share dynamic range, one column-wise factor serves both. To switch precision without altering traversal, each nonzero carries three value streams (\texttt{FP64-val}, \texttt{FP32-val}, \texttt{BF16-val}) over the same structural indices, so precision selection changes only which stream is read. Columns are padded to a multiple of two with sentinel-marked zeros, preserving vectorized access and structural alignment without affecting logical sparsity. EvoSpMV thus runs heterogeneous tile precisions in one indexing framework with no kernel-switch overhead, providing the stable backbone EvoRun embeds into the CG critical path.

\subsubsection{Cross-Level Balance Tuning}

The row-block height $R$ is EvoSpMV's main preprocessing knob for cross-level balance. \sysname tunes $R$ over $\{32,64,\dots,16384\}$ to even non-empty tile counts across row-blocks for inter-tile warp-group balance, which in turn reshapes the nnz distribution within 64-column tiles for lane-level balance.


\subsection{EvoRun: Low-Overhead In-Loop Runtime}
\label{sec:low_overhead_control}

As illustrated in Fig.~\ref{fig:EvoRun}, EvoRun is the runtime module of \sysname. It integrates EvoCtrl and EvoSpMV directly into the CG iteration loop. Policy inference remains part of the critical path, but its cost is reduced and amortized through a fused device-side inference kernel and interval-based action reuse. The remaining CG operators are fused and replayed through a CUDA Graph, reducing launch and intermediate-memory overhead around the controlled SpMV.

\paragraph{Policy Inference}
Since policy execution runs inside CG iteration, its cost must stay minimal. We fuse the whole policy inference into one kernel, avoiding intermediate write-backs and extra launches. Policy weights are quantized offline to FP16 and reordered into a WMMA-friendly layout, enabling FP16 Tensor Core inference with FP32 accumulation.

\paragraph{Interval Reuse Scheduling}
Two observations motivate reusing precision decisions across iterations. First, column-block precision patterns exhibit strong temporal persistence between adjacent iterations, so per-iteration policy invocation is largely redundant. Second, precision transitions are asymmetric: delaying an \emph{upward} transition may leave a block insufficiently accurate and even break convergence, whereas delaying a \emph{downward} one only costs efficiency. Together, these suggest that the policy can be invoked periodically, provided upward transitions are not unduly delayed.

Guided by this, we set the reuse interval $K^\star$ offline. From a per-iteration run of \sysname, we record each column block's precision and, over candidate intervals $1, 2, 4, 8, \dots$, measure the fraction of upward transitions that reuse would delay. $K^\star$ is the largest interval keeping this fraction below 0.1. Online, the policy runs every $K^\star$ iterations, reusing its decision within each window.

\paragraph{Fused Execution and Graph Replay}
To complete the runtime path, EvoRun compresses the remaining CG operators---dot-product reductions, residual update, and search-direction update---into a single cooperative fused-step kernel that completes global reduction, scalar broadcast, and vector updates entirely on device via grid-wide synchronization, removing repeated launches, redundant memory traffic, and intermediate write-backs. The \texttt{SpMV + fused-step} backbone is further captured in a CUDA Graph to eliminate multi-kernel launch overhead, while policy inference stays outside the graph and triggers only once every $K^\star$ iterations.

\section{Experiments}

\subsection{Experimental Setup}

\begin{figure*}
    \centering
    \includegraphics[width=1.0\linewidth]{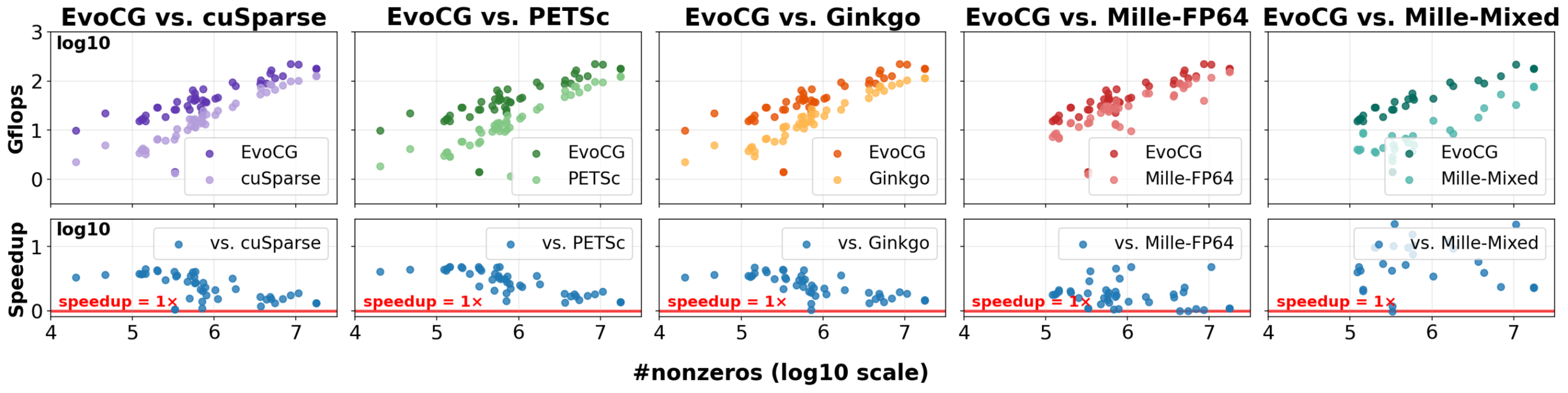}
    \caption{End-to-end performance vs.\ cuSPARSE, PETSc, Ginkgo, and two Mille-feuille variants across 47 SPD matrices: achieved GFLOPS (top) and solver-level speedup of \sysname over each baseline \rev{on a log$_{10}$ axis} (bottom), against nonzeros (log$_{10}$).}
    \label{fig:end2end_perf}
\end{figure*}

\paragraph{Hardware and software}
All experiments run on an NVIDIA A100 (40\,GB HBM) with an AMD EPYC 7K62 CPU (8 cores, 16 threads) and 93\,GiB host memory, under Ubuntu 22.04, CUDA 12.2 (driver 535.216.01), GCC 11.4.0, CMake 3.22.1.

\paragraph{Baselines}
We compare EvoCG against five baseline configurations from four solver systems: PETSc v3.24.5~\cite{mills2021toward} , Ginkgo v1.11.0~\cite{anzt2022ginkgo}, an FP64 CG solver based on cuSPARSE~\cite{cusparse} and cuBLAS, and two configurations of Mille-feuille~\cite{yang2024mille}. The cuSPARSE baseline uses CSR SpMV from cuSPARSE-cu12 v12.3.1 together with cuBLAS for the remaining vector operations. We denote the two Mille-feuille configurations as Mille-FP64 and Mille-Mixed. Mille-FP64 executes the entire solver in FP64, whereas Mille-Mixed uses the released mixed-precision implementation. PETSc, Ginkgo, and the cuSPARSE baseline do not support mixed-precision CG in our experimental setting and therefore run entirely in FP64.

\paragraph{Precision configuration}
The tile-level action space includes FP64, FP32, and BF16. FP8 is excluded because A100 offers no practically beneficial FP8 path for our workflow, and FP8 actions destabilize RL training and prevent reliable policy convergence.

\paragraph{Unified solver configuration}
All methods use the same configuration: initial guess $x_0=0$, right-hand side $b=A\mathbf{1}$ (exact solution $\mathbf{1}$), the recursively updated relative-residual stopping criterion $\|r_k\|_2/\|r_0\|_2\le 10^{-10}$, and a maximum of 10,000 iterations.

\paragraph{Datasets}
We evaluate \sysname on 47 sparse matrices from the SuiteSparse Matrix Collection~\cite{davis2011university}, covering eight scientific domains. Our benchmark set includes all SuiteSparse matrices that meet three criteria: they are symmetric positive definite (SPD), have dimension greater than 10{,}000, and converge within 10{,}000 CG iterations. \rev{The dimension threshold ensures GPU utilization and SpMV dominance. The iteration threshold keeps solver-in-the-loop training feasible. The set was \emph{not} filtered by conditioning and includes difficult instances such as \texttt{nd6k} and \texttt{cbuckle} (estimated condition numbers $>10^7$). }

\paragraph{Training configuration}
The precision-control policy is trained with PPO using $\gamma=0.99$, learning rate $10^{-4}$, 10 epochs per update, minibatch size 128, clip ratio 0.2, value-loss coefficient 0.25, entropy coefficient 0.02, update interval 1024, hidden sizes $[64,64]$, gradient clipping 0.5, advantage standardization, and GAE parameter $\lambda=0.95$. \rev{These hyperparameters follow stability-oriented defaults~\cite{schulman2017ppo} refined by a single small validation sweep, shared unchanged across all matrices and never tuned per matrix.}

\rev{\paragraph{Training protocol}
\sysname-Mixed uses a matrix-specific policy trained offline on the target matrix. This setting targets repeated-solve deployments, where training is a one-time cost amortized over subsequent solves. Sec.~\ref{subsection:generalization} evaluates transferring one \texttt{mesh2e1}-trained policy to unseen matrices. 
Reported times cover the online solve only, excluding offline policy training, whose cost is discussed in Sec.~\ref{subsection:cost_analysis}.}

\subsection{End-to-End Solver Performance}
\label{subsec:e2ePerformance}

As shown in Fig.~\ref{fig:end2end_perf}, under one common configuration and convergence criterion, \sysname is consistently the fastest solver across the 47 SPD matrices, with geometric-mean speedups of 2.52$\times$, 3.06$\times$, 2.48$\times$, 1.93$\times$, and 6.99$\times$ over cuSPARSE, PETSc, Ginkgo, Mille-FP64, and Mille-Mixed, respectively. The corresponding maximum speedups reach 4.47$\times$, 4.89$\times$, 4.44$\times$, 4.87$\times$, and 22.49$\times$. For Mille-FP64 and Mille-Mixed, we report results only on matrices where their released implementations complete successfully.

Against cuSPARSE, PETSc, and Ginkgo, the gain comes from full-stack co-design: adaptive mixed precision cuts arithmetic cost whenever the Krylov state tolerates inexactness, Tile-CSC and the warp-centric kernel absorb the execution irregularity of sparsity plus precision heterogeneity, and fused execution removes redundant launches and intermediate memory traffic. Crucially, none of these gains is erased by convergence degradation.

The Mille-FP64 comparison is particularly informative: Mille-FP64 often outperforms cuSPARSE, PETSc, and Ginkgo thanks to its fused-kernel design, yet it remains a fixed-precision FP64 solver without \sysname's co-design around precision heterogeneity, so \sysname achieves a clear 1.93$\times$ geometric-mean speedup.
\sysname achieves an even larger speedup over Mille-Mixed, and the reasons are instructive. \rev{The mega-kernel design of Mille-Mixed introduces device-side spin-wait synchronization, four precision copies, per-iteration casts, and \texttt{d\_vis}-guided warp branches. \sysname avoids these overheads with a staged pipeline, in which Tile-CSC mixed-precision SpMV is followed by a cooperative fused kernel synchronized grid-wide via \texttt{cooperative\_groups}. This contrast illustrates a broader point: mixed precision brings not only lower arithmetic cost but also execution heterogeneity, since tiles of different precision incur different amounts of work. Mixed precision thus pays off only when the storage format, kernel mapping, and solver execution are co-designed to absorb this imbalance.}

\subsection{Ablation of Dynamic Precision Control}
\label{subsection:ablation_precision}

\begin{figure}[htb]
    \centering
    \includegraphics[width=1.0\linewidth]{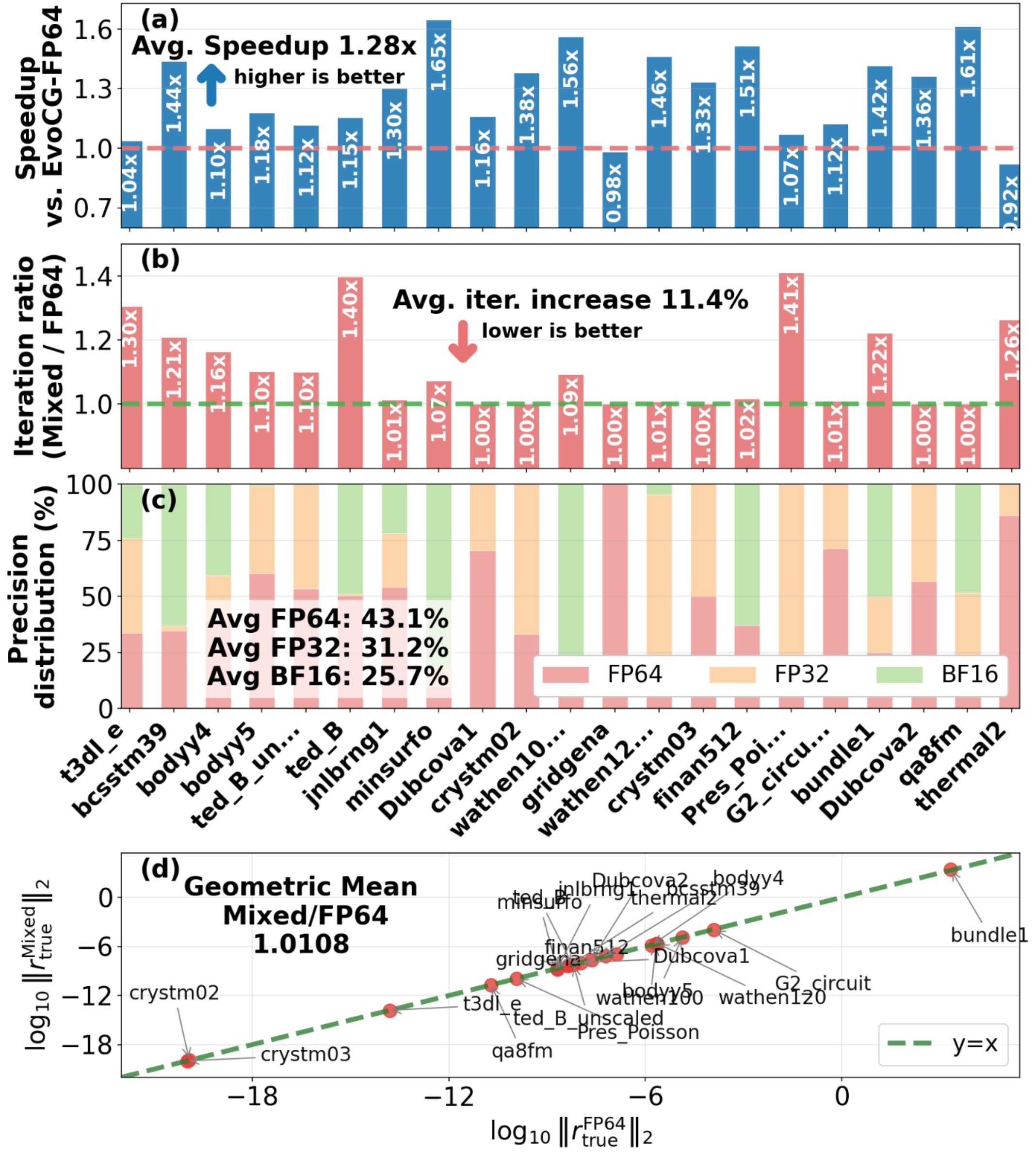}
    \caption{Mixed-precision ablation on CG against EvoCG-FP64, the double-precision solver on the identical execution substrate. (a) Speedup of EvoCG-Mixed over EvoCG-FP64. (b) Increase in iterations (Mixed vs.\ FP64). (c) Precision distribution per matrix. (d) Comparison of the final FP64-recomputed true-residual norms of EvoCG-Mixed and EvoCG-FP64 on a $\log_{10}$ scale. The dashed line denotes $y=x$.}
    \label{fig:ablation_mixed}
\end{figure}

We next isolate dynamic precision control. \rev{The reference \sysname-FP64 is our CG solver on the \emph{identical} Tile-CSC substrate, fused kernels, and CUDA-Graph runtime, with every tile in FP64. It produces exactly the same iteration counts and final residuals as cuSPARSE-, PETSc-, and Ginkgo-based FP64 CG, so the comparison isolates the precision controller from the execution engine.}

Fig.~\ref{fig:ablation_mixed} shows that the learned controller improves time-to-solution while preserving FP64-level convergence quality. For readability, Fig.~\ref{fig:ablation_mixed} visualizes 21 representative cases, but all aggregate values below are computed over the full set of 47 matrices. Fig.~\ref{fig:ablation_mixed} (a) reports an geometric-mean speedup of $1.28\times$ over all 47 matrices, while showing 21 representative cases for readability; Fig.~\ref{fig:ablation_mixed} (b) shows that this speedup is achieved with only an $11.4\%$ average increase in iteration count; and Fig.~\ref{fig:ablation_mixed} (c) shows substantial variation in precision mix across matrices, indicating matrix-dependent decisions rather than a fixed global policy, while more than 56\% of the SpMV computation is executed in low precision on average. 
\rev{For numerical validation, after each solve we recompute the true residual $r_{\mathrm{true}}=b-Ax$ in FP64 for both variants. Fig.~\ref{fig:ablation_mixed}(d) plots $\log_{10}\|r_{\mathrm{true}}^{\mathrm{Mixed}}\|_2$ against $\log_{10}\|r_{\mathrm{true}}^{\mathrm{FP64}}\|_2$: the points closely track the $y=x$ line, with a geometric-mean Mixed/FP64 true-residual norm ratio of only $1.0108$, confirming that EvoCG-Mixed reaches nearly the same final accuracy as EvoCG-FP64. The true-residual recomputation is performed outside the timed solver execution.}

Fig.~\ref{fig:ablation_mixed}(a)--(c) show that the benefit of dynamic precision control is dominated by a favorable group of 13 matrices, including \texttt{jnlbrng1}, \texttt{minsurfo}, and \texttt{crystm02}. On this set, mixed precision improves performance by $36.5\%$ with only a $3.2\%$ increase in iteration count, so the reduced SpMV cost translates almost directly into solver-level speedup. 
A second group, including \texttt{t3dl\_e}, \texttt{bodyy4}, \texttt{ted\_B}, and \texttt{Pres\_Poisson}, incurs a larger iteration increase of $31.8\%$ on average but still delivers a $9\%$ speedup, since the larger low-precision fraction outweighs the longer recurrence: iteration growth does not negate the benefit as long as per-iteration savings exceed the convergence penalty. Boundary cases are rare. \texttt{thermal2} uses little low precision (\(86\%\) FP64 and \(14\%\) FP32) yet still incurs a \(26\%\) iteration increase, leading to an \(8\%\) slowdown, while \texttt{gridgena} remains entirely in FP64 and thus exposes only the policy overhead, causing a modest \(2\%\) slowdown.

\rev{Overall, dynamic precision control is robustly beneficial, and notably, no \sysname-Mixed run diverges or misses the tolerance on any of the 47 matrices. In the worst cases, the policy autonomously retreats to a predominantly FP64 schedule, bounding the downside to the mild slowdowns above, roughly the policy overhead.}

\subsection{Analysis of Convergence Stability}
\label{subsection:convergence_stability}

\begin{figure}[ht]
    \centering
    \includegraphics[width=1\linewidth]{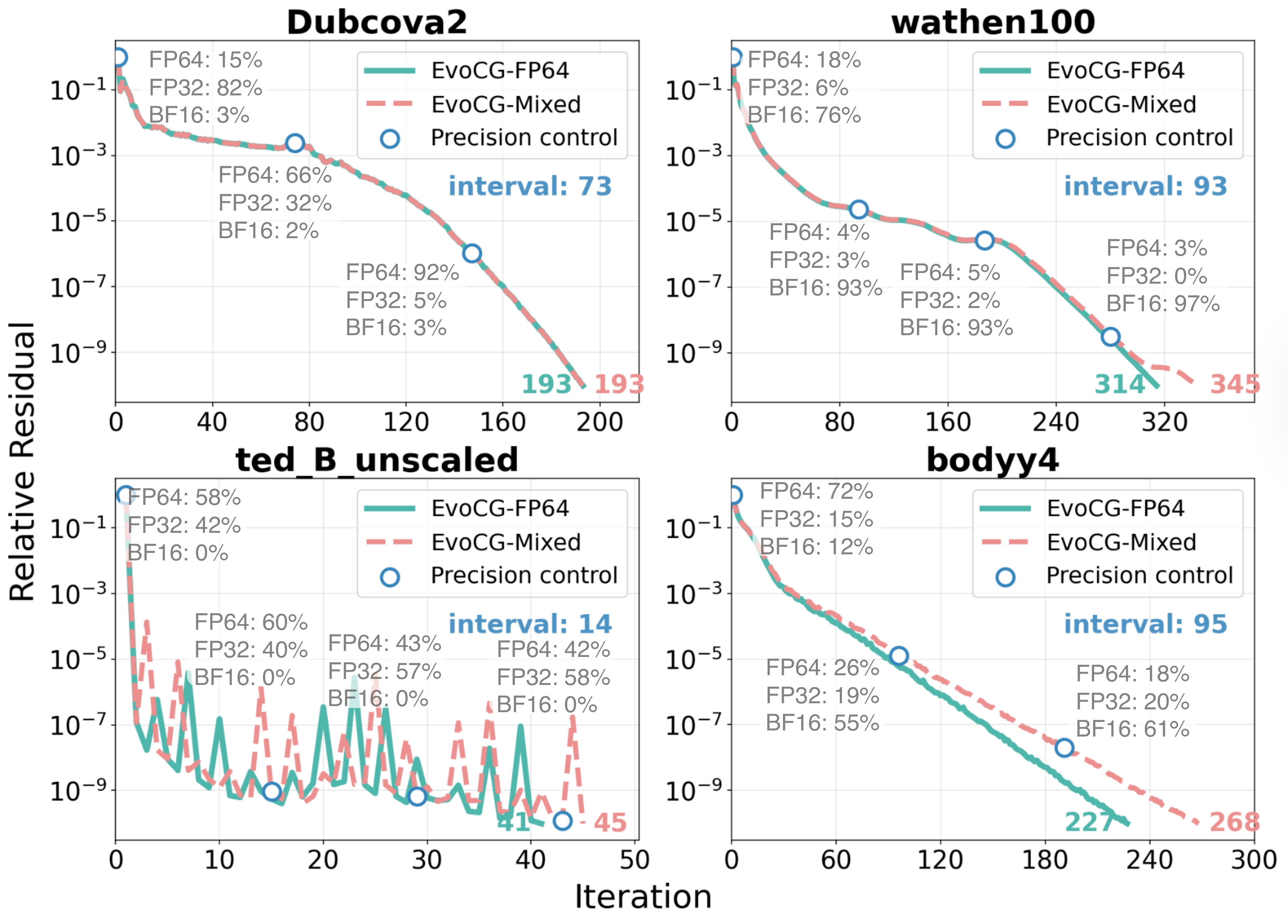}
    \caption{Trajectories of the recursively updated relative residual for \sysname-FP64 and \sysname-Mixed on four representative matrices.}
    \label{fig:convergence}
\end{figure}

A central question is whether \sysname can exploit low precision while preserving CG's essential convergence dynamics. Fig.~\ref{fig:convergence} shows that EvoCG reaches the target tolerance and closely tracks the FP64 trajectory of the recursively updated residual.

\begin{figure*}[t]
    \centering
    \includegraphics[width=1\linewidth]{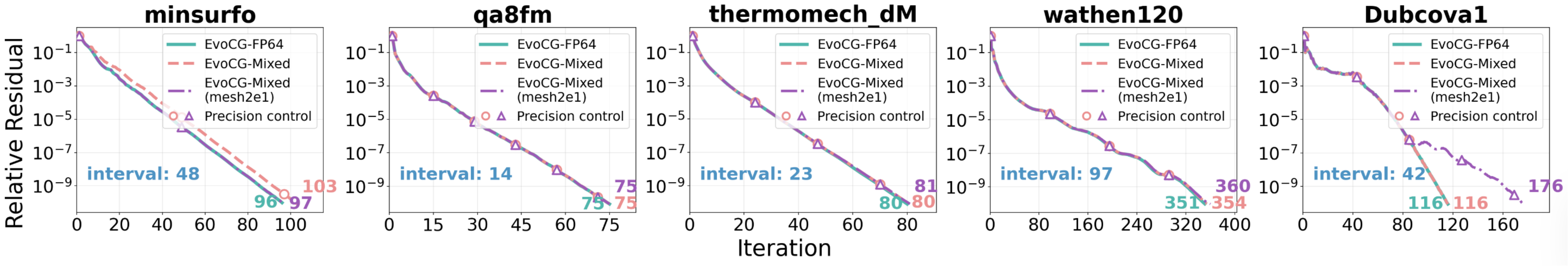}
    \caption{Representative convergent cases of cross-scale policy transfer from \texttt{mesh2e1} to larger unseen matrices.}
    \label{fig:generalization}
\end{figure*}

The first two cases, \texttt{Dubcova2} and \texttt{wathen100}, show that \sysname-Mixed can be nearly convergence-neutral. \texttt{Dubcova2} converges in the same number of iterations as \sysname-FP64, and \texttt{wathen100} stays tightly aligned with only a moderate increase from 314 to 345 iterations even though BF16 dominates most later intervals. Both descend smoothly without late-stage stagnation or oscillation, so \sysname can exploit low precision aggressively on tolerant problems at little or no penalty.

The other two cases represent a more sensitive regime. On \texttt{ted\_B\_unscaled}, both solvers show the sharp non-monotone fluctuations typical of this problem, yet the mixed-precision run reaches the tolerance with only a small delay, 45 iterations versus 41. On \texttt{bodyy4} the gap is larger, 268 versus 227, but the residual still decreases steadily. The penalty thus appears as reduced convergence efficiency rather than stagnation, breakdown, or loss of numerical control.

The precision trajectories further show that the controller does not follow a fixed monotone schedule: \texttt{wathen100}, \texttt{bodyy4} and \texttt{ted\_B\_unscaled} shift toward more BF16 in later stages, whereas \texttt{Dubcova2} grows more conservative near the end. \sysname thus adapts to the numerical tolerance of each solve rather than simply raising precision over time.

\subsection{Cross-Scale Transfer of the Learned Precision Policy}
\label{subsection:generalization}

\rev{Using $p_k$ as the policy state ties the learned policy to numerical distributions rather than to matrix-specific sparsity patterns, raising the possibility that a policy trained on one problem transfers to others. To test this,} we train the controller only on \texttt{mesh2e1}, a tiny $306 \times 306$ matrix with 2018 nonzeros and under 15 minutes of training, then transfer it to all 47 unseen matrices without retraining. \rev{We do not claim universal transfer. The transferred policy outperforms EvoCG-FP64 on 23 of the 47 matrices. Across these 23 improved cases, it achieves a geometric-mean speedup of 1.21$\times$. Among the remaining matrices, 15 converge without a performance gain and 9 fail to converge. Transfer is thus real across scales but bounded by problem family, since $p_k$ encodes numerical distributions shared within a family rather than across all matrices. We treat it as a zero-retraining fast path for problems near the training family, with matrix-specific training as the default and fallback whenever transfer fails.}

\rev{Fig.~\ref{fig:generalization} reports five representative matrices spanning outcomes from better to worse than the matrix-specific policy: \sysname-Mixed is the matrix-specific policy, \sysname-Mixed (\texttt{mesh2e1}) the transferred policy, and \sysname-FP64 the double-precision reference.} On \texttt{minsurfo}, the transferred policy even improves over the matrix-specific one, cutting iterations from 103 to 97. On \texttt{qa8fm} the two are identical at 75 iterations, and on \texttt{thermomech\_dM} they differ by only one iteration. On \texttt{wathen120}, transfer costs only a minor 354-to-360 increase. \texttt{Dubcova1} is the only clear outlier, needing 176 iterations instead of 116. 



\rev{
\subsection{Extension to Jacobi-Preconditioned CG}
\label{subsection:pcg}

We further evaluate whether a policy trained on unpreconditioned CG remains effective when the solver dynamics are altered by preconditioning. We apply \sysname to Jacobi-PCG on all 47 matrices without retraining, reusing the policy trained for the corresponding unpreconditioned system. The policy continues to control only the SpMV $q_k=Ap_k$, while the Jacobi application $z_k=M^{-1}r_k$ and all scalar operations remain in FP64. Both \sysname-Mixed and \sysname-FP64 converge on all 47 matrices. \sysname-Mixed achieves a $4.17\times$ geometric-mean speedup over cuSPARSE-based Jacobi-PCG and a $1.15\times$ speedup over \sysname-FP64, with only a $4.99\%$ average increase in iteration count. The latter comparison isolates the benefit of adaptive precision on the same preconditioned execution substrate, showing that the policies learned for CG transfer effectively to Jacobi-PCG without retraining.
}

\rev{
\subsection{Why Learned Precision Control?}
\label{subsec:whyLearnedPC}

We next ask whether EvoCtrl is necessary, or whether a simple rule-based scheduler can achieve comparable robustness and performance. To isolate the precision-control mechanism, we replace only EvoCtrl with a residual-threshold rule adapted from Mille-Mixed~\cite{yang2024mille}, while keeping the same Tile-CSC layout, mixed-precision SpMV kernel, and EvoRun runtime. Given the convergence tolerance $\epsilon=10^{-10}$, the rule assigns each tile a precision from the largest entry magnitude $m_i=\max_j |p_k^{(i)}[j]|$ in its segment: BF16 if $m_i<\epsilon\times10^{-1}$, FP32 if $m_i\in[\epsilon\times10^{-1},\epsilon)$, and FP64 otherwise. Thus, the comparison attributes differences in convergence and execution time directly to the precision decisions.

\begin{table}[h]
\centering
\caption{Comparison of EvoCG, residual-threshold precision scheduling,
and cuSPARSE FP64.}
\label{tab:rule_based_comparison}

\footnotesize
\setlength{\tabcolsep}{2.2pt}
\renewcommand{\arraystretch}{1.06}

\begin{tabular}{@{}lrrrrrr@{}}
\toprule
& \multicolumn{3}{c}{Iterations}
& \multicolumn{3}{c}{Time (ms)} \\
\cmidrule(lr){2-4}
\cmidrule(lr){5-7}
Matrix
& \multicolumn{1}{c}{EvoCG}
& \multicolumn{1}{c}{Rule}
& \multicolumn{1}{c}{FP64}
& \multicolumn{1}{c}{EvoCG}
& \multicolumn{1}{c}{Rule}
& \multicolumn{1}{c}{FP64} \\
\midrule

finan512
& 58 \inc{1}
& 57
& 57
& \textbf{2.68}
& 4.06
& 7.39 \\

bcsstm39
& 499 \inc{84}
& 415
& 415
& \textbf{12.76}
& 18.37
& 46.71 \\

cvxbqp1
& 5095 \inc{14}
& 5083 \inc{2}
& 5081
& \textbf{161.05}
& 281.30
& 583.76 \\

jnlbrng1
& 156 \inc{2}
& 154
& 154
& \textbf{4.35}
& 5.65
& 19.00 \\

bundle1
& 264 \inc{49}
& 216 \inc{1}
& 215
& \textbf{11.50}
& 15.98
& 27.11 \\

ted\_B
& 682 \inc{189}
& 821 \inc{328}
& 493
& \textbf{13.48}
& 15.52
& 51.10 \\

ted\_B\_unscaled
& 45 \inc{4}
& 41
& 41
& \textbf{1.13}
& 1.28
& 5.05 \\

cbuckle
& 4849 \inc{7}
& 4845 \inc{3}
& 4842
& \textbf{240.54}
& 261.22
& 524.91 \\

Dubcova2
& 193
& 10000
& 193
& \textbf{11.88}
& 1571.47
& 25.51 \\

Pres\_Poisson
& 3428 \inc{993}
& 10000
& 2435
& \textbf{131.05}
& 560.07
& 266.85 \\

\bottomrule
\end{tabular}

\raggedright
\scriptsize
\end{table}

The rule exposes two undesirable operating regimes. It either retains mostly FP64 computation and provides limited speedup, or uses low precision aggressively and risks substantial iteration growth or convergence failure. Under the same 10,000-iteration limit, the rule converges on only 27 of the 47 matrices, whereas EvoCG converges on all 47. Table~\ref{tab:rule_based_comparison} reports representative matrices covering both regimes. The conservative regime dominates the convergent cases shown in the table: on \texttt{finan512}, \texttt{bcsstm39}, \texttt{cvxbqp1}, \texttt{jnlbrng1}, \texttt{bundle1}, \texttt{ted\_B\_unscaled}, and \texttt{cbuckle}, the rule keeps most tiles in FP64, yielding iteration counts close to FP64 but leaving a substantial performance gap to EvoCG. In contrast, the aggressive regime appears on \texttt{ted\_B}, where the rule requires 821 iterations compared with 493 for FP64, and more broadly in the 20 matrices where the rule fails to converge within the 10,000-iteration limit. The table shows two representative non-convergent cases, \texttt{Dubcova2} and \texttt{Pres\_Poisson}. These results show that entry magnitude alone does not reliably capture tile-level numerical sensitivity.

EvoCtrl optimizes time-to-solution rather than iteration count. Across the eight displayed matrices on which both methods converge, EvoCG is $1.36\times$ faster than the rule on geometric-mean and is faster in every case. This remains true when EvoCG executes more iterations. On \texttt{bcsstm39}, for example, EvoCG requires 499 iterations instead of 415, yet reduces execution time from 18.37~ms to 12.76~ms. 

Overall, a scalar magnitude threshold cannot simultaneously expose substantial low-precision opportunities and preserve robust convergence across matrices. EvoCtrl instead learns state-dependent, tile-wise precision decisions from the evolving Krylov process, allowing EvoCG to reduce per-iteration cost without entering the unstable regime of an overly aggressive fixed rule.
}


\subsection{Cost Analysis}
\label{subsection:cost_analysis}

While \sysname improves solver performance through adaptive mixed-precision execution, it also incurs three extra costs: online policy execution, auxiliary Tile-CSC and precision metadata storage, and offline policy training. We analyze them separately to clarify the system overhead.

\begin{figure}[b]
    \centering
    \includegraphics[width=1\linewidth]{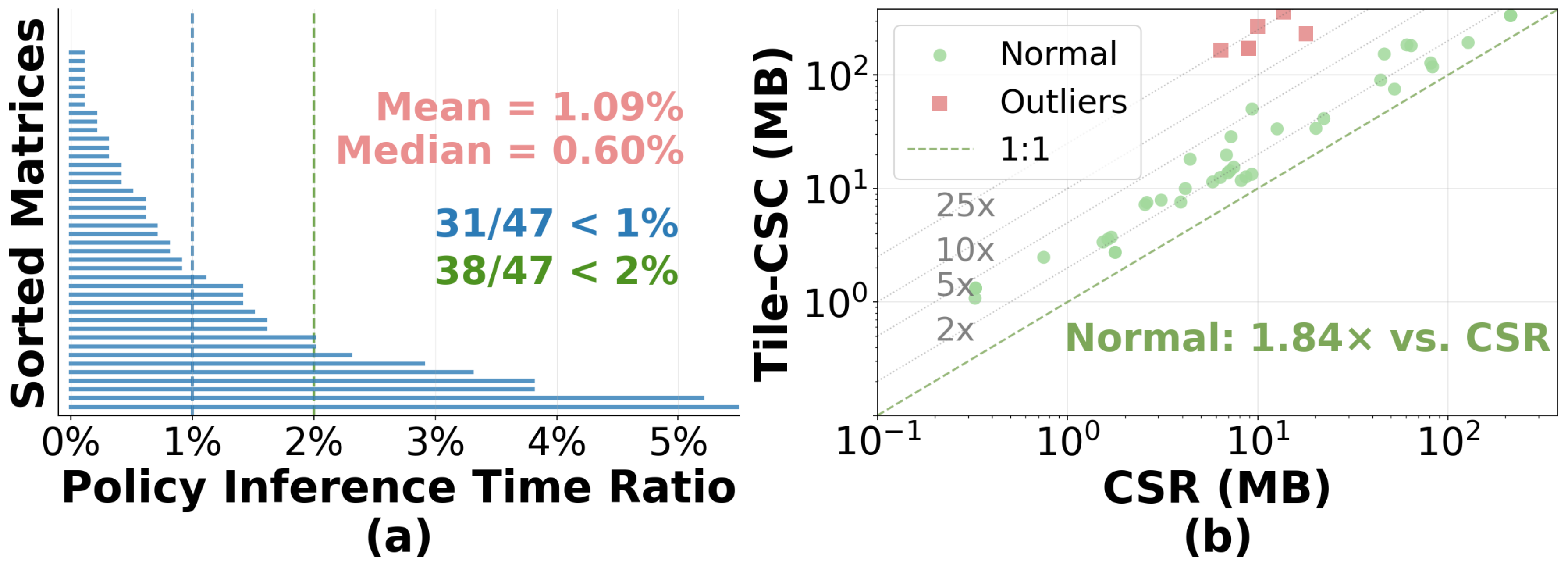}
    \caption{Cost analysis of EvoCG. (a) Policy inference time ratio across matrices. (b) Memory usage of Tile-CSC relative to CSR.}
    \label{fig:cost_analysis}
\end{figure}

\paragraph{Policy cost}
Fig.~\ref{fig:cost_analysis}(a) reports the runtime overhead of embedding the learned policy into the CG loop. Across all 47 matrices, policy inference accounts for only 1.09\% of end-to-end execution time on average, with a median of 0.60\%. Moreover, 31 matrices remain below 1\% overhead, and 38 remain below 2\%. This result confirms that the lightweight two-layer MLP, fused device-side inference kernel, and profile-guided interval $K^\star$ together make neural precision control a practical in-loop mechanism rather than a solver-level bottleneck.

\paragraph{Memory cost}
The memory overhead of Tile-CSC remains well controlled for the dominant portion of the workload set. As illustrated in Fig.~\ref{fig:cost_analysis} (b), on the 40 non-outlier matrices, Tile-CSC requires only 1.84$\times$ the memory of CSR overall. In this normal regime, the extra storage mainly comes from two deliberate choices: maintaining pre-quantized low-precision backups alongside the FP64 values, and using padding to regularize tile layout for warp-centric execution. This memory increase is necessary for the unified precision path of \sysname in Sec.~\ref{subsubsection:unified_precision_path}. By materializing multiple value streams offline, \sysname can switch precision at runtime through direct value-stream selection, without incurring online quantization overhead or changing the traversal path. This design is important for preserving execution regularity and fully exposing the benefit of low-precision computation. A small number of outliers incur substantially higher memory overhead because their nonzeros are dispersed across many tiles, which magnifies tile-level structural and padding cost. Overall, Tile-CSC pays a controlled and necessary memory premium in exchange for efficient unified-kernel mixed-precision execution.

\paragraph{Training cost}
The training cost of \sysname is entirely offline and amortized across future solves. In our experiments, the policy converges within 1000 episodes. Under this setting, training on a small matrix such as \texttt{mesh2e1} takes less than 15 minutes. As discussed in Sec.~\ref{subsection:generalization}, the mesh2e1-trained policy provides useful but non-universal cross-scale transfer, reducing the need for matrix-specific training on a subset of related problems. Matrix-specific training typically takes several hours and is therefore intended for workloads that repeatedly solve the same or closely related systems.

\section{Related Work}

Recent work accelerates CG and related Krylov solvers via tile-grained single-kernel GPU execution~\cite{yang2024mille}, adaptive mixed-precision PCG~\cite{guo2025adaptive}, RL-based precision tuning~\cite{chen2025mixed}, and architectural support~\cite{song2023callipepla}. Broader mixed-precision Krylov methods explore inexact operators, residual correction, robust preconditioning, and nested solvers~\cite{yamazaki2022mixed,ina2023new,suzuki2025nested}. These studies establish the value of precision heterogeneity but treat it as an execution option, a safeguard, or an external tuning layer. \sysname instead makes precision an online control variable derived from the solver state and consumed by tile-grained sparse execution in the loop. \rev{Table~\ref{tab:positioning} summarizes this positioning.}

\rev{
\begin{table}[h]
\centering
\caption{Positioning of \sysname relative to the closest mixed-precision CG systems.}
\label{tab:positioning}
\small
\renewcommand{\arraystretch}{0.95}
\setlength{\tabcolsep}{3pt}
\resizebox{\columnwidth}{!}{%
\begin{tabular}{lccccc}
\toprule
 & Decision & Granularity & \shortstack{Closed\\loop} & \shortstack{Format\\co-design} & \shortstack{In-loop\\overhead} \\
\midrule
Mille-mixed~\cite{yang2024mille} & static rule & tile & \ding{55} & \checkmark & \ding{55} \\
Callipepla~\cite{song2023callipepla} & fixed scheme & stream & \ding{55} & FPGA ISA & \ding{55} \\
RL tuning~\cite{chen2025mixed} & RL (offline) & operator & \ding{55} & \ding{55} & \ding{55} \\
Adaptive PCG~\cite{guo2025adaptive} & numerical rule & operator & partial & \ding{55} & \ding{55} \\
\sysname & RL (in-loop) & tile & \checkmark & \checkmark & \checkmark \\
\bottomrule
\end{tabular}%
}
\end{table}
}

\enlargethispage{4\baselineskip}
\sysname also differs from the broader CG-acceleration literature. Prior work reduces communication, synchronization, and launch overheads through pipelining, communication-avoiding variants, and kernel fusion~\cite{rupp2016pipelined,aliaga2015systematic,trotter2025cpu,carson2018adaptive,carson2018stability,solomonik2014tradeoffs}. Other work improves the sparse substrate through format redesign, vectorization, or load balancing~\cite{liu2015csr5,merrill2016merge,cheshmi2022vectorizing,anzt2020load}.
Mixed-precision numerical linear algebra likewise trades arithmetic fidelity for performance under proper numerical control~\cite{van2004inexact,clark2010solving,abdelfattah2021survey,higham2022mixed}. \sysname closes this loop, making precision selection a runtime solver action.

\section{Conclusion}

This paper presents \sysname, a mixed-precision CG solver that treats tile-wise precision as a closed-loop control problem over the evolving Krylov state. By co-designing EvoCtrl, EvoSpMV, and EvoRun, \sysname\ learns when reduced precision is numerically tolerable and translates it into end-to-end speedup.
Across 47 SPD matrices from eight scientific domains, \sysname\ delivers the best overall end-to-end performance among all evaluated baselines \rev{and extends to Jacobi-PCG without retraining}.
More broadly, \sysname\ suggests a path toward self-evolving HPC solvers that regulate their numerical behavior on modern low-precision hardware.

\section*{Acknowledgements}
This work was supported by the National Key Research and Development Program of China under Grant 2024YFB4505202, the National Natural Science Foundation of China under Grant 62632004, and the Hubei Provincial Natural Science Foundation of China under Grant 2026AFA002.

\newpage

\bibliographystyle{ACM-Reference-Format}
\bibliography{ref}

\end{document}